\documentclass[reprint, amsmath,amssymb, aps,10pt]{revtex4-1}

\usepackage{graphicx}
\usepackage{dcolumn}
\usepackage{bm}
\usepackage{subfigure}
\usepackage{color}

\newcommand{\be}{\begin{equation}}
\newcommand{\ee}{\end{equation}}
\newcommand{\bea}{\begin{eqnarray}}
\newcommand{\eea}{\end{eqnarray}}
\newcommand{\ba}{\begin{array}}
\newcommand{\ea}{\end{array}}

\begin{document} 
\title{Self-energy of strongly interacting Fermions in  Medium: a Holographic Approach}
\author{Yunseok Seo$^{\dagger}$, Sang-Jin~Sin$^{\dagger}$ and Yang~Zhou$^{\dagger *}$}
\address{
$\dagger$ Department of Physics, Hanyang University,
Seoul 133-791, Korea\\
$*$ Center for Quantum
Spacetime, Sogang University, Seoul 121-742, Korea}
 

\begin{abstract}
 We consider the  self-energy of strongly interacting   fermions
 in the medium using gauge/gravity duality of $D4/D8$ system. 
 We  study the  mass generation of the thermal and/or dense  medium   
 and  the collective excitation called plasmino, by considering the spectral function of fermion and its dispersion relation. 
Our results are very different from  those of the hard thermal loop method: 
  for zero density, there is no thermal mass or  plasmino in any phase. 
  Plasmino in deconfined phase is not allowed in $D4/D8$ set up.  
  In the confined phase, there is plasmino modes only for  a window of  density.
 \end{abstract}

\maketitle
\section{   Introduction}
The study of fermion self-energy in medium has a long history due to its fundamental importance in studying electronic as well as nuclear matter system. When the excitations are strongly interacting, perturbative field theory method cannot give a reliable result since 
the diagrams should be truncated to the   ladder or rainbow types, 
which can not be justified in strong coupling. 
Furthermore in the presence of chemical potential, the lattice technique is not much useful due to the 
sign problem. Therefore it is  worthwhile to utilize the gauge/gravity duality for this tantalizing problem. 
The gauge/gravity duality  was used to study the   fuzzy fermi surface~\cite{Lee:2008xf}   and the non-fermi liquid nature ~\cite{Liu:2009dm,Faulkner:2009wj,Cubrovic:2009ye} of the strongly interacting system, for recent developments, we refer to~\cite{Bhattacharya:2012we,Hartnoll:2010gu,Sachdev:2011ze,Allais:2012ye}.

The weakly interacting field theory (QED or QCD ) results for the fermion self energy in medium  can be summarized by the existence of the plasmino mode  \cite{Braaten:1990it} and thermal mass generation of order $gT$. 
Plasmino is a collective mode whose dispersion curve has a minimum at finite momentum. 
However, recent study of the thermal field theory \cite{Harada:2007gg,Nakkagawa:2011ci},  by solving   Schwinger-Dyson equation numerically, showed that 
the thermal mass is reduced as the coupling grows. It raises an interesting questions what happens to the thermal mass and more generally  to the hard and soft momentum scales and magnetic mass scale, $T, gT, g^2T$ respectively,   in the strong coupling limit. 

In this letter we  study  the dispersion relation  and thermal mass generation using  gauge/gravity duality and    report  a feature of plasmino in   strongly coupled system.   
We consider  $D4/D8$ model \cite{Sakai:2004cn} and turn on a fermion field on the flavor brane world-volume  with finite quark/baryon number density \cite{Kim:2006gp, Nakamura:2006xk, Bergman:2007wp}. 
In deconfined phase, 
the fermions represent the fundamental strings, connecting the horizon of black branes and the probe $D8$.
In the zero 't Hooft coupling limit,  it is a bi-fundamental representation.
In the large 't Hooft coupling limit,     
$D4$-branes disappear and the color index  of the fermions should disappear also. 
In confined phase, to describe a baryon 
we need to introduce Witten's baryon vertex \cite{Witten:1998xy}.  $N_c$ strings are connecting it to the probe $D8/\bar{D8}$. 
The dynamics of connecting  open  strings  define that of the baryon vertex as well as the $D8$.  
Since each  string  gives fermionic mode, total vibrational dynamics should be described by the composite operator 
 coming from the product of all $N_c$ fermions. 
 This is the baryon in confined phase which is fermion if $N_c$ is odd. 
 We replace this composite operator by a massive fermion field. 
So 
our construction for baryon is rather phenomenological. 
We consider just one flavor brane case mostly. 
Previously  fermions on a probe brane in adjoint representation, called mesino,  were studied in \cite{Heise:2007rp,Ammon:2010pg}, which are different from ours. 

By solving the Dirac equations in each phase, 
 we obtain dispersion relations for the fermionic  excitations in medium. 
 Our results show that for zero density, there is  
  no thermal mass generation and no plasmino in any phase,   
  which  is sharply different from weakly coupled field theory result. 
 To get  plasmino in deconfined phase we need to add  quark mass as well as quark density.
  In $D4/D8$ set up, the quark mass is not allowed so plasmino is forbidden in deconfined phase. 
   In the confined phase, there is a plasmino mode only for a certain window of density.
   
\section{Plasmino  in field theory}
We begin our discussion by giving a brief discussion of fermionic collective excitations in a plasma. The fermion propagator is written as
\be
G(p)= \frac1{\gamma\cdot p-m-\Sigma(p)}\ ,\ee
 where $\Sigma=\gamma_\mu \Sigma^\mu$ is self-energy. 
The gauge invariant result is available in the hard (high temperature) thermal loop approximation (HTL) where fermion mass $m$ can be ignored since it is small compared with $T$ or $\mu$.
The most noticeable effect of the medium is effective mass generation.  The HTL result of effective mass is given by ~\cite{Sigma}
$ m_f^2={1\over 8}g^2C_F\left(T^2+{\mu^2\over\pi^2}\right)\ ,$ where $C_F=1$ for electron and $C_F=4/3$ for quark.  
There are two branches of dispersion curves $\omega=\omega_\pm (p)$ whose asymptotic forms are given 
by 
\bea
p<<m_f &:& \quad \omega_\pm(p) \simeq m_f \pm \frac13 p \label{asym}\ , \\ 
p>>m_f &:& \quad \omega_\pm (p) \simeq p\ .
\eea
$\omega_+$ is the normal branch and $\omega_-$ is the one describing plasmino that 
has been extensively investigated
\cite{Braaten:1990it,Sigma,Pisarski:1993rf,Braaten:1991dd,Blaizot:1996pu,Braaten,Schaefer:1998wd,Weldon:1989ys,Pisarski:1989wb, Baym:1992eu,Peshier:1998dy,Blaizot:1993bb,Peshier:1999dt,Mustafa:2002pb}. 
 For a review we refer to~\cite{Bellac, Kapusta}. The presence of plasmino is  important since it may enhance production rate of  the light di-lepton ~\cite{Peshier:1999dt}. 

In HTL approximation, fermions can be regarded as massless 
therefore two branches  $\omega_\pm$ can be characterized by chirality and helicity. The ratios of chirality and helicity for $\omega_\pm$ are $\pm 1$ respectively.
In our work, we do not neglect the fermion mass  so we consider the presence of plasmino as the presence of minimum of dispersion curve 
at finite momentum. We will conclude that the plasmino exists if the dispersion relation has a minimum at finite momentum.

Notice that  plasmino in HTL approximation exists in high temperature whatever the density is.
However, in the strongly coupled limit, we will show that plasmino disappears at zero density in deconfined phase. And it can survive only for a certain window of  chemical potential in confined medium.
Especially, at zero chemical potential, there is no plasmino. 
We find that  the constant value $1/3$ in  Eq.(\ref{asym}) will be replaced by a function of density. See figure 7.  

\section{Holographic set up}
\vskip0.2cm
Let us  now set up a holographic model to calculate the fermion self-energy, first in confined phase. 
We use Sakai-Sugimoto (SS) model~\cite{Sakai:2004cn}  where    probe $D8/\bar{D}8$ is embedded
in  black $D4$ branes background. 
To introduce  finite density, we turn on $U(1)$ gauge field on the probe brane. The sources of the $U(1)$ gauge field are end points of strings which are emanating from   horizon  in deconfined phase and from  baryon vertex in confined phase.  

{\it Confined Phase}~~
The geometry of confined phase with Euclidean signature is given by 
\bea
 ds^2 &=& \left({U\over
R}\right)^{3/2}\left(\delta_{\mu\nu} dx^\mu dx^\nu + f(U)dx_4^2\right) \cr
&+& \left({R\over U}\right)^{3/2}\left({dU^2\over f(U)} + U^2
d\Omega_4^2\right)\ , \label{eq8}
\eea 
where both the time and the Kaluza-Klein direction are periodic:  $x^0\sim x^0+\delta x^0$, $x_4\sim
x_4+\delta x_4$ and 
$
 f(U) = 1-\frac{U_{KK}^3}{U^3}\ ,~~ U_{KK} =
\left({4\pi\over 3}\right)^2 {R^3\over \delta x_4^2}\ .
$
Here following the original  Sakai-Sugimoto model, we consider only trivial embedding of the flavor eight brane where $D8$ and $\bar{D8}$ brane are located at the antipodal position in $x_4$ direction. 
The dynamics of probe branes with $U(1)$ gauge field on it  can be
governed by the DBI action as follow
\bea 
S_{D8} &=& -T_8 \int d^9 x e^{-\phi} \sqrt{\det(g_{MN} + 2\pi \alpha' F_{MN})}\cr
&=&{\cal N}  \int dr~ r^4 \left[r^{-3}\left({1/ f(r)} - a'_0(r)^2\right)\right]^{1/2},
\eea
where 
$ {\cal N}={T_8\Omega_4 v_3\delta x^0 R^5/g_s}\ ,$ and $\Omega_4$ is the unit four sphere volume and $v_3$ is the three Euclidean space volume. Notice that in the case of antipodal configuration of $D8/\bar{D8}$, the embedding becomes trivial i.e. $x_4'(r)=0$.
For later convenience, we use dimensionless quantities 
\be
 r = U/R\ ,~r_0=U_{KK}/R\ ,~a_0 = 2\pi \alpha' A_0/R.
 \ee
From the equation of motion for $U(1)$ gauge field, we get conserved dimensionless quantity
${ra_0^\prime(r)\over \sqrt{r^{-3}\left({1/ f(r)} - a_0'(r)^2\right)}} = D $ ,
where $D$ is an integral constant representing the baryon density.

Chemical potential can be defined as the value of $a_0(r)$ at the infinity, once proper IR boundary condition is imposed at $r =r_0$ .  
In confined case, the baryon vertex which is $D4$ brane wrapping on $S^4$ with $N_C$ fundamental strings plays the role of source of the $U(1)$ gauge filed.
The energy of the source should be included in the total free energy of the system and contributes to the chemical potential also.

Accordingly we set $a_0(r_0)=m/q$ to include the baryon mass $m$  in the chemical potential $\mu$: 
 \be\label{totalmu}
 \mu=m/q+\mu_0=m/q+\int_{r_0}^\infty a'_0 dr.
 \ee

 In confined phase, $\mu_0$ is the contribution only from the gauge potential and
 $m$ is identified as baryon mass, which  can be calculated from DBI action of the baryon vertex.
 It  is related  to the 5 dimensional Lagrangian mass $m_5$ as we will see later.
 Notice that such identification is not true in bottom up  AdS model, where bulk fermion mass should be related to the  conformal dimension of an operator. 
 
 {\it Deconfined Phase}~~
The deconfined geometry is given by
 double Wick rotating $x_4$ and time from the confined one. 
The DBI action for probe brane can be written as
\be
S_{D8}={\cal N}' \int dr \,\,r^4 \sqrt{r^{-3}(1-a_0'(r)^2)}.
\ee
The gauge field profile on eight brane is solved as before,
${ra_0^\prime(r)\over \sqrt{r^{-3}\left(1- a_0'(r)^2\right)}} = D' $,
where $D'$ is a constant.
 In deconfined phase the end points of fundamental strings play the role of sources of the $U(1)$ gauge filed.
Since the probe branes always fall into the black hole horizon, the fermion representing quark is massless  in this phase.

The IR boundary condition of $a_0$ field determined by the regularity condition at the horizon is given by
$
a_0(r_H)=0\ ,
$ where $r_H=U_H/R\ .$
This condition is also consistent with the argument in previous section. In deconfined case, mass of source is zero hence it does not contribute to the IR boundary condition for chemical potential.  
 
\begin{figure}
\begin{center}
\subfigure[]{\includegraphics[angle=0,width=0.23\textwidth]{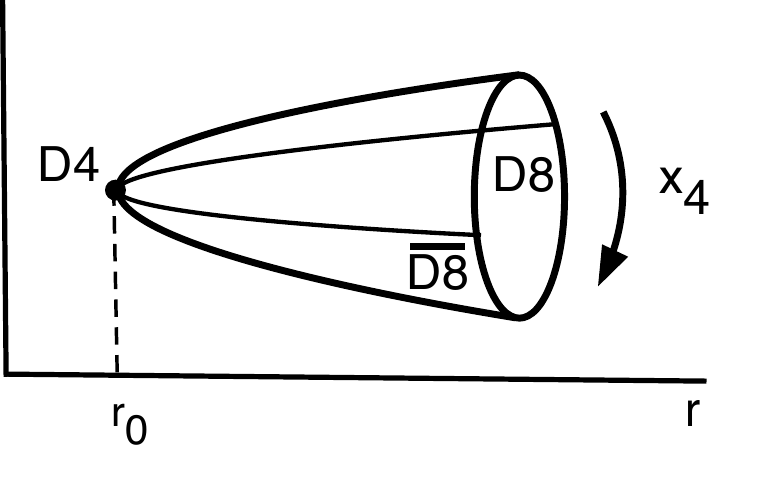}}
\subfigure[]{\includegraphics[angle=0,width=0.23\textwidth]{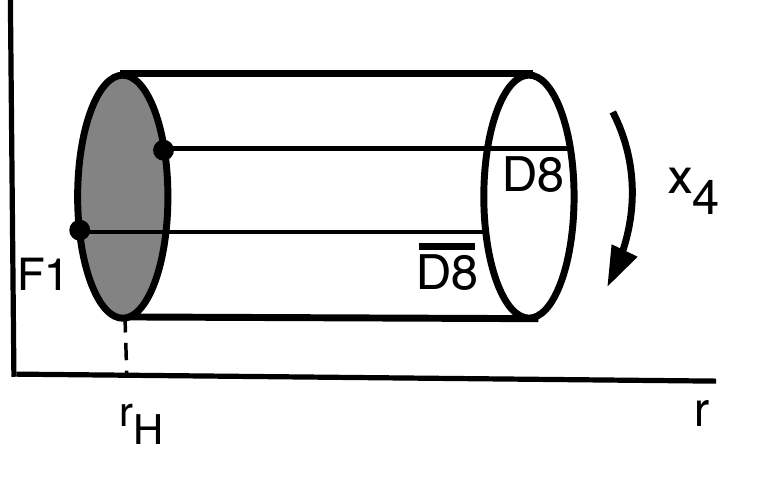}}
\caption{(a) Embedding of $D8/\bar{D8}$ brane in confined phase. $D4$ baryon vertices locate at $r=r_0$. (b) Embedding of probe branes in deconfined phase. The source of $U(1)$ gauge field is located on the black hole horizon.  }
\label{fig:SSemb}
\end{center}
\end{figure}

 The schematic configuration of probe branes are drawn in Figure \ref{fig:SSemb}.

 \section{  Fermionic Spectrum}
Now we study the fermionic spectral function in the holographic dual system. We put a probe fermion field in the world volume of flavor $D8$ brane in the Sakai-Sugimoto model.
We ignore the $S^4$ part and work in the effective 5 dimensional world volume following the original work of Sakai-Sugimoto model. The induced 
metric  is written as $ ds_5^2 = g_{\mu\nu} dx^\mu dx^\nu + g_{rr}dr^2\ .$
We use the minimal action \be S =
\int d^5x \sqrt{-g}\  i \left(\bar\psi\Gamma^M D_M\psi -
m_5 \bar\psi\psi\right)\ , \ee where the covariant derivative is $ D_M = \partial_M + {1\over 4} \omega_{abM} \Gamma^{ab} - i q A_M\ .$ Here $M$ denotes the bulk spacetime index while $a,b$ denote the tangent space index. After a factorizing  
$
\psi=(-gg^{rr})^{-1/4}e^{-i \omega t + i k_i x^i}\Psi\ ,
$ the Dirac equation for $\Psi$ can be give by
\be\label{dirac1}
\sqrt{g_{ii}/ g_{rr}}(\Gamma^{\underline{r}}\partial _r - m_5\sqrt{g_{rr}})\Psi + i K_\mu \Gamma^{\underline{\mu}}\Psi = 0\ ,
\ee where $ K_\mu = (-v(r), k_i)\ ,~ v(r)=  (\omega + qa_0)\sqrt{-g_{ii}/g_{tt}}\ .$
Following the procedure in~\cite{Faulkner:2009wj},
we rewrite the Dirac equation (\ref{dirac1}) in terms of  two component spinors
\be\label{EOMM}
(\partial_r + m_5 \sqrt{g_{rr}}\sigma^3)\Phi_\alpha = \sqrt{g_{rr}/ g_{ii}}(i\sigma^2 v(r) +(-1)^\alpha k\sigma^1)\Phi_\alpha\ ,
\ee where $\sigma^i$ are Pauli matrices and $\alpha=1,2$ denoting the up and down two spinor respectively.
Further decomposing $ \Phi_1=(y_1, z_1)^{T}\ , \Phi_2=(y_2, z_2)^{T}\  , $ we get equations of motion 
for $y_\alpha$ and $z_\alpha$. When $\alpha=2$ we have
\begin{eqnarray}
(\partial_r + m_5 \sqrt{g_{rr}})y_2(r) &=& \sqrt{g_{rr}/ g_{ii}}(+ v(r)+k)z_2(r)\ ~\label{wave1}\\
(\partial_r - m_5 \sqrt{g_{rr}})z_2(r) &=& \sqrt{g_{rr}/g_{ii}}(-v(r)+k)y_2(r)\ .~\label{wave2}
\end{eqnarray}
By replacing $k$ by $-k$, we obtain the equations of motion for $y_1$ and $ z_1$. We would like to define the following new variables
 $ G_1(r) : = {y_1(r)/z_1(r)}\ ,
 G_2 (r) := {y_2(r)/ z_2(r)}\ . $
 The retarded Green function in boundary field theory is obtained as 
 \be
 G^R_{1,2} = \lim_{\epsilon\rightarrow 0} e^{-8 m_5Rr^{1/4}} G_{1,2}(r)|_{r=1/\epsilon}\ ,
 \ee where $G_1$ and $G_2$ satisfy the following equations
  \begin{eqnarray}
\sqrt{g_{ii}\over g_{rr}}&&\partial_r G_\alpha + 2m_5\sqrt{g_{ii}}G_\alpha \cr 
&&=(-1)^\alpha k +
v(r)+\left((-1)^{\alpha -1} k+ v(r)\right)G_{\alpha}^2\ . ~~\label{floweq2}
\end{eqnarray}
Now the remain task is to solve  (\ref{floweq2}) by imposing proper boundary conditions. 

{\it Confined phase}~~
In the confined phase of Sakai-Sugimoto model, $v(r)$ function appearing in the Dirac equation is given by
$ v(r)= \omega + q a_0(r)\ .$
where the $U(1)$ flux solution on flavor brane is obtained from DBI action as
  \be
a_0(r)= \mu + \int_{\infty}^{r} d\hat r \left({f(\hat r)^{-1}D^{2} \over \hat r^5 +D^{2} 
}\right)^{1/2} \ .
  \ee
  Notice that $g_{rr} = R^2 r^{-3/2}f(r)^{-1}$ which diverges at $r_0$. The boundary conditions for flow equation (\ref{floweq2}) can be found by requiring (\ref{floweq2}) regular at $r=r_0$. They are given by
   \be\label{bcondition1}
G_\alpha(r_0) =  {-m  R  +\sqrt{m^2 R^2 +k^2- {\hat \omega}^2}\over (-1)^\alpha k- \hat\omega}\ ,
\ee 
where $ \hat\omega=\omega+m\ .$ And we define a 4 dimensional vacuum mass $m:=m_5\sqrt{g_{ii}}|_{r=r_0}\ .$
Notice that imposing the boundary condition for retarded (advanced) Green function corresponds to $\omega\rightarrow\omega + i \epsilon$ ($\omega\rightarrow\omega - i \epsilon$).

{\it Deconfined phase}~~
In the deconfined phase, $v(r)$ function is given by 
$ v(r)= {\omega + q a_0(r)\over \sqrt{f}}$ and 
the electric flux is given by \be
a_0(r)= \mu + \int_{\infty}^{r} d\hat r \left({D'^{2} \over \hat r^5 +D'^{2} 
}\right)^{1/2}\ . \ee Due to the black hole horizon, we imposing the in-falling boundary condition at $r=r_H$.
The boundary condition at $r=r_H$ is obtained as 
\be \label{bcondition2}
G_{1,2}(r_H) = i\ . \ee
The  fermion dispersion relation
 can be obtained  by searching for poles of spectral function, which is the imaginary part of retarded Green function.
We solve (\ref{floweq2}) numerically with  IR boundary conditions
 (\ref{bcondition1}) in confined phase and (\ref{bcondition2}) in deconfined phase.  
\begin{figure}
\begin{center}
\subfigure[]{\includegraphics[angle=0,width=0.23\textwidth]{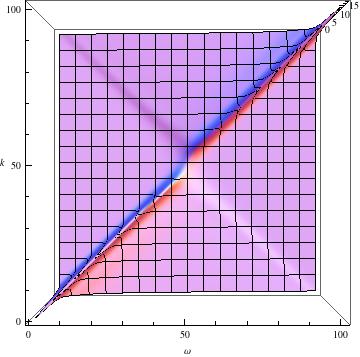}}
\subfigure[]{\includegraphics[angle=0,width=0.23\textwidth]{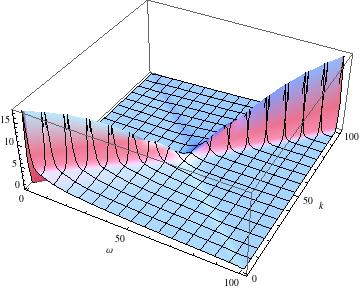}}
\caption{Spectral functions of $G^R_2$ at zero density: the dispersion curve passes through the origin and thermal mass vanishes.   Top view and side view of 3D plot of $\rm{Im} G^R_2$ with $T=1,\ m=0$. True range of both $\omega$ and $k$ is $[-5,5]$. }
\label{zerodensity}
\end{center}
\end{figure}


\section{Numerical Results}
Now we discuss the results of fermionic spectral function. First in the deconfined phase, the self-energy term gets some  imaginary part due to the in-falling boundary condition. 
If the imaginary part is not large we can read off the dispersion relation from the positions of the maxima of spectral function. 
The 3D plot of spectral function with {\it zero density} is shown in Figure \ref{zerodensity}.
  Since the fermions here are deconfined quarks, 
  we set $m=0$. Our main question  here is whether 
  thermal mass can be generated by finite temperature effect in strong coupling limit. In Figure \ref{zerodensity}
the dispersion curve passes through the origin and this feature is  independent of temperature although it is illustrated  for $T=1$. As a result, no thermal mass is generated and  there is no plasmino in deconfined phase with zero density. The absence of thermal mass is actually one of most drastic difference compared with the weakly coupled field theory result. Namely 
   \bea
      m_T &=&\frac1{\sqrt{6}} g T \; \hbox{in weak coupling},  \cr
       m_T &=& 0 \;\;\;\;\;\; \hbox{ in strong coupling}. \eea
  The result of vanishing thermal mass is actually consistent with 
  a recent claim made in \cite{Nakkagawa:2011ci} by numerical study of Dyson-Schiwinger equation in the strong coupling region.

 If we turn on finite density in the system, density effect can generate effective mass. 
 However for massless fermion we have an exact Green function $G_R=i$ at $k=0$, so it is impossible to observe peak structure near $k=0$. We can not observe plasmino mode in finite density for the massless fermion.

What happen if we added a finite bulk fermion mass for curiosity? 
We find that density effect can  generate 
effective mass as well as  plasmino mode for large enough chemical potential.
\begin{figure}[ht!]
\begin{center}
\includegraphics[angle=0,width=0.35\textwidth]{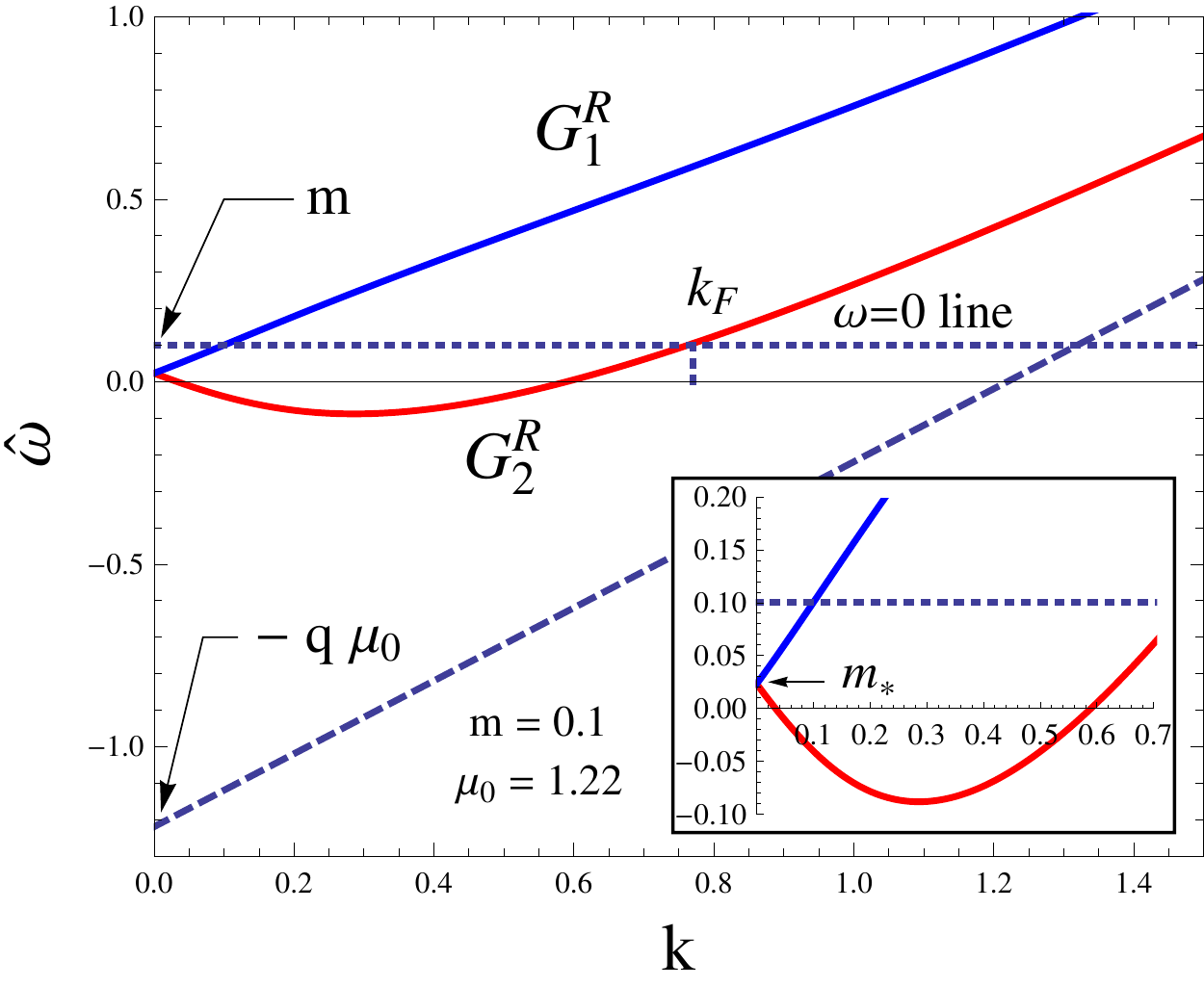}
\caption{Dispersion relations in confined phase. The upper and lower branches describe the normal fermion $\omega_+$ and plasmino $\omega_-$ respectively. Dotted line denotes light cone.
\label{PlasminoDis}}
\end{center}
\end{figure}

\begin{figure}[ht!]
\begin{center}
\includegraphics[angle=0,width=0.3\textwidth]{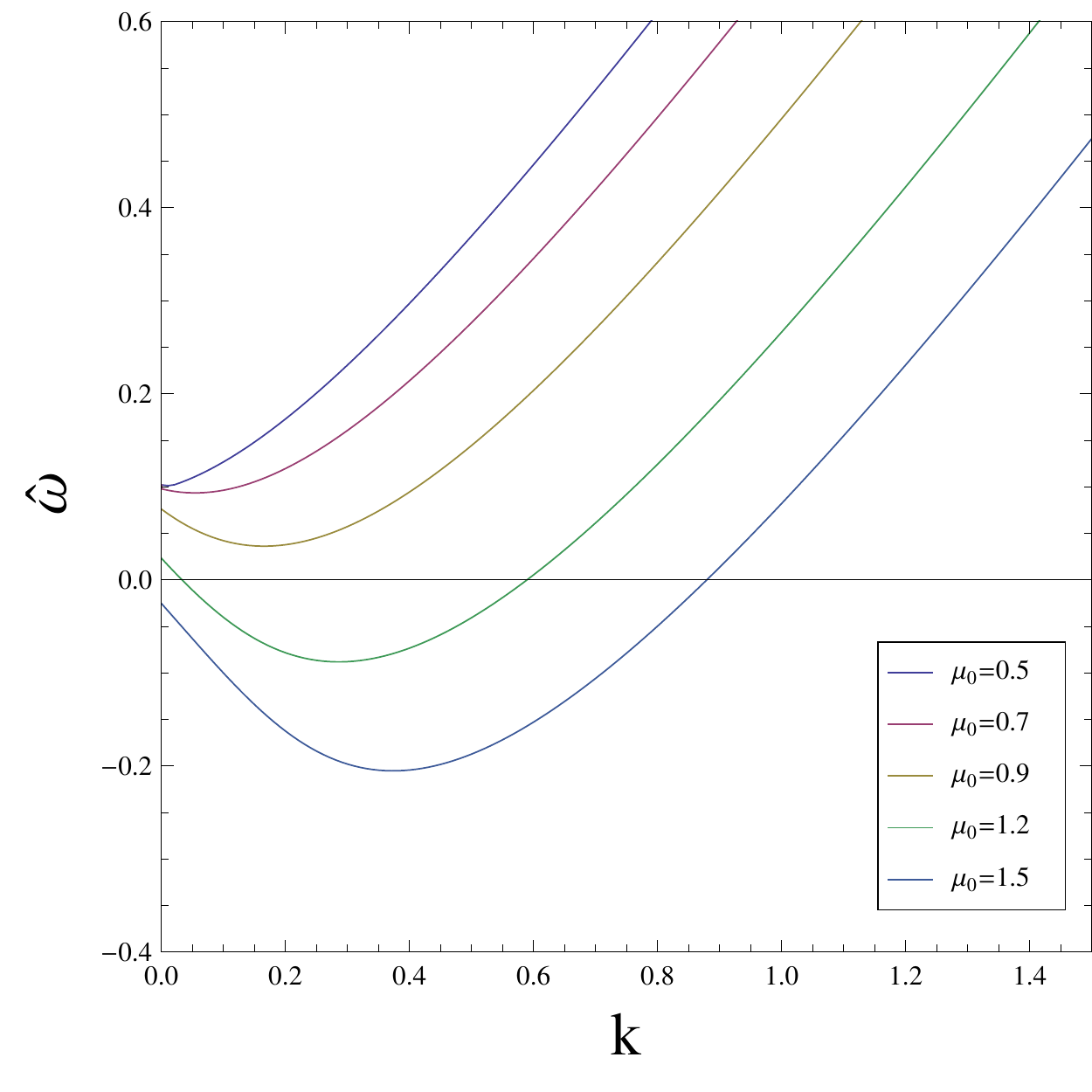}
\caption{Dispersion relations of $G^R_2$ in confined phase. As chemical potential increases, dispersion curve moves down. Parameter $m=0.1$. \label{PlasminoDis1}}
\end{center}
\end{figure}

\begin{figure}[ht!]
\begin{center}
{\includegraphics[angle=0,width=0.3\textwidth]{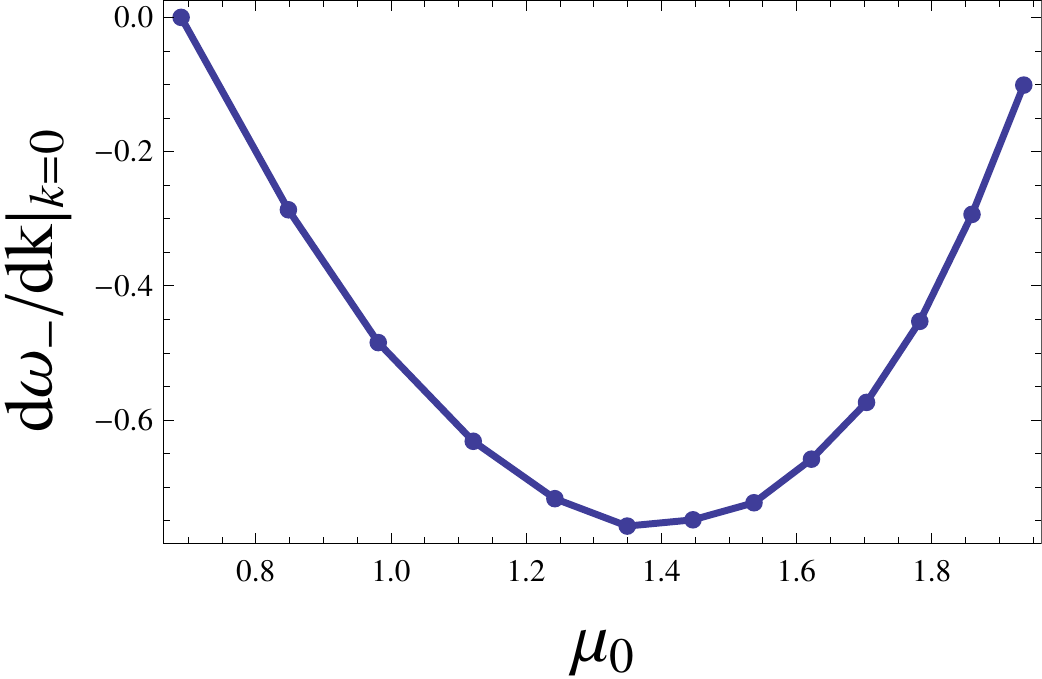}}
 \caption{  $\mu_0$ dependence of $\alpha=\frac{d\omega}{dk}$ at $k=0$. The curve is plotted only in the density window  $\mu_1 \le \mu_0 \le \mu_2$ where there is plasmino. Parameter $m=0.$. Notice that $\alpha=\frac{d\omega}{dk}$ at $k=0$ has a  constant value -1/3 for the 
 weakly coupled field theory.  }
\label{fig3}
\end{center}
\end{figure}
Now we turn to the confined phase. In this phase, all the peaks of spectral function are delta function-like peaks.
In general the Fermi momentum is defined by $\omega(k_F)=0\ .$
We plot dispersion relations $\omega=\omega_\pm(k)$ in  Figure \ref{PlasminoDis}. 

{\it Observation of plasmino}~~
 We observe a plasmino dispersion relation characterized by the presence of the minimal energy at finite momentum. 
As we change chemical potential, the slope of $\omega_-$ at $k=0$ changes. We plot the slope $\alpha(\mu_0)$ as a function of density in figure \ref{fig3}. In  HTL approximation, the value of slope at $k=0$ is  $ -\frac13$  independent of  density or temperature.

The high density behavior of the dispersion curve is complex and we will report it elsewhere.  
We restrict ourselves to the density range
where traditional plasmino mode exists.
We could observe plasmino  only in a  chemical potential window 
$\mu_1<\mu_0<\mu_2 $.
 When $m=0.1$, the window is given by $\mu_1=0.69\ ,\ \mu_2=1.94$. As $m$ increases, this window gets wider. Inside the window, as density becomes larger, dispersion curve moves down and bends more and more as shown in Figure \ref{PlasminoDis1}.
This may be compared with field theory result in weak coupling, where effective  mass and plasmino are generated for any density.
 
\begin{figure}[ht!]
\begin{center}
\subfigure[]{\includegraphics[angle=0,width=0.23\textwidth]{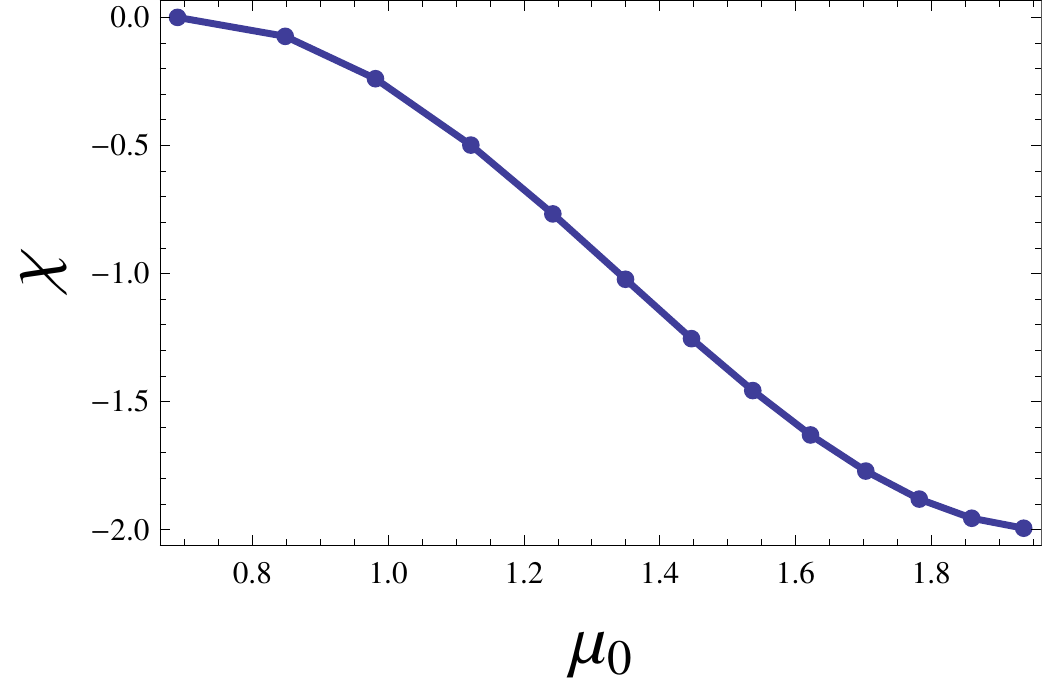}}
\subfigure[]{\includegraphics[angle=0,width=0.23\textwidth]{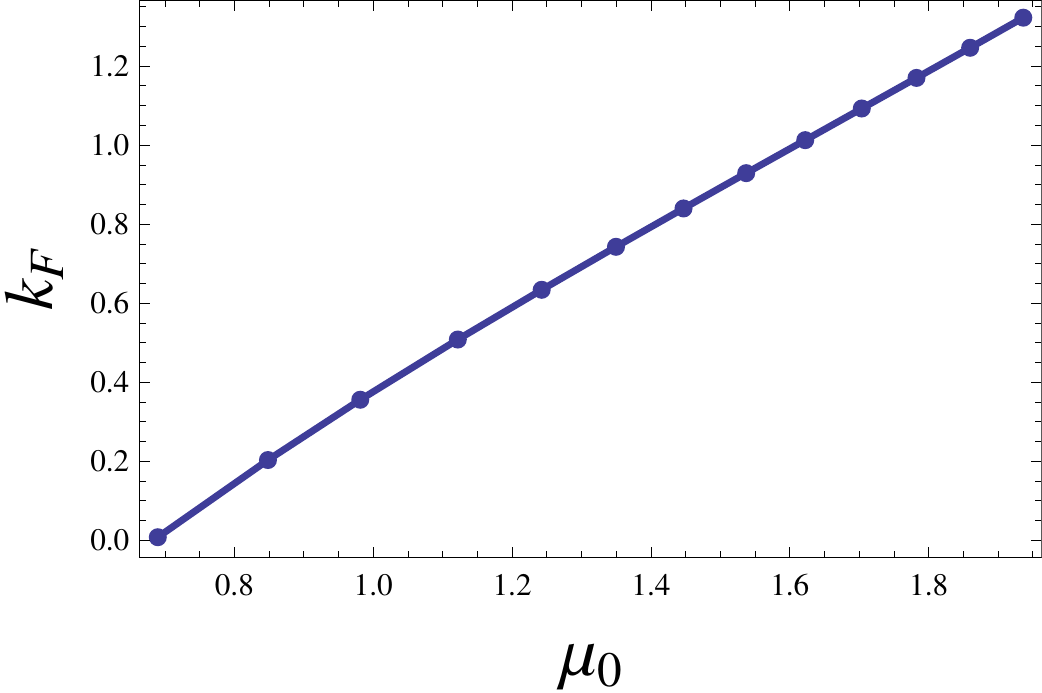}}
\caption{ a): $\mu_0$ dependence of $\chi$. 
 b): $\mu_0$ dependence of $k_F$. Parameter $m=0.1$.}
\label{Fig4}
\end{center}
\end{figure}

 We can read off   baryon mass in  medium defined by $m_*:=\hat \omega(0)$.   
We will show $m_*$  is a monotonically increasing function of chemical potential.   
 The mass shift in medium is defined as $\delta m :=m_*-m$. The normalized mass shift  quantity $ \chi(\mu_0):=\delta m /m$ is plotted in figure \ref{Fig4}.
 Notice that  for massless case, there is no mass correction at all.   

If we turn off both charge and bulk mass in confined phase, we have exact solution 
$
 G_2(r)= \sqrt{k+\omega\over k-\omega}\ ,$
 which is independent of radial direction.  
 
{\it Fitting dispersion relation}~~
 Now we try to fit the dispersion curve of plasmino. We write the Green function as
%
\be \label{G2}
G^R_2(\omega, k) =\frac{Z}{\omega -\delta m - \frac{k(k-C)}{k+B}}\ .
\ee Although the numerator $Z$ is some undetermined function, the pole information is assumed to be contained only in the denominator. Parameter $B$ and $C$ can be written in terms of $\delta m=m_*-m$, $k_F$ and $v_F$ as follows
\be
B=\frac{k_F^2 (1-v_F)}{\delta m +k_F v_F},~~~~~C=\frac{\delta m^2 +2\delta m k_F +k_F^2 v_F}{\delta m +k_F v_F}\ ,
\ee 
where $\delta m$ is mass shift, $k_F$ is Fermi momentum and $v_F$ is Fermi velocity. $v_F$ is defined as the slope at $k=k_F$.
Expanding  (\ref{G2}) near Fermi momentum, we get
\be
G^R_2(\omega, k) \sim \frac{Z}{\omega -v_F (k-k_F) -\Sigma},
\ee
where the self-energy near Fermi momentum is obtained as
\be
\Sigma = \delta m + \frac{(1-v_F)(\delta m +k_F v_F)}{k_F (\delta m +k_F)} \cdot (k-k_F)^2 +{\cal O}((k-k_F)^3).
\ee
The vanishing mass shift limit $\delta m=0$ corresponds to the massless fermion in confined phase. In this limit, self-energy becomes
\be
\Sigma = \frac{(1-v_F) v_F}{k_F } \cdot (k-k_F)^2 +{\cal O}((k-k_F)^3).
\ee
 
\section{ Conclusion and Discussion} 
In this paper we  discuss  our observations  on the characters of fermion's  self-energy  in the dense medium. By using gauge/gravity dual, we recover  normal  and plasmino branch. 
In deconfined phase, we showed that  for zero density, there is no thermal mass generation.   
 In the weakly coupled thermal gauge theory, $T$,  $gT$  and  $g^2T$ provide  three  well separated scales, 
 with physical interpretations: hard momentum, thermal mass for fermion(or electric screening mass for gluon), magnetic mass  respectively. These masses play the role of the infrared regulator of different scales.  
 Such separation does not happen for large coupling   $\sim O(1)$, and we believe that we can not define any such physical scale either for strong coupling. 
In the confined phase,  plasmino excitations are present only  for a  window of the density. The group velocity of the plasmino mode at zero momentum  turns out to be density dependent rather than a constant, $-1/3$, showing the deviation of the HTL approximation. 
It is also worthwhile to notice that without medium effect, there is no mass correction. 
We found a simple empirical formula for plasmino dispersion relations.
We expect that the phenomena we discovered here is common in many holographic models, so that 
it is a universal character. 

{
There is an issue
whether $m_5$ can be calculated in a way following \cite{Ammon:2010pg}, 
 where one starts with  massless fermion in 10 dimension and performs the KK reduction from 10 dim  to 5 dim.
 The answer is  negative   because D4 background does not have a direct product structure. 
 To see this, let us consider \be (\gamma.D_x+  \gamma.D_y -m_{10})\Psi(x,y)=0\ ,\ee
 where $x$ represents the  non-comact directions and $y$  does  the compact directions.
 For D3 background, where manifold is direct product $AdS_5\times S^5$,
 the $\gamma.D_y$ term  produces  $\sim (l+2)/R_{ads}$ which can be interpreted as a mass term.  
 While in our model, D4 background is not a direct product. Therefore the KK reduction in our case generates  
term $\sim (l+2)/r$  which is a  potential rather than a   mass. 
One may call  such potential term as a ``mass function" which depends on the radius.  We will study the effect of mass function in future work.}

{\it Note added}: In the closing period of this work, an observation on two branches of dispersion relation appeared in~\cite{Herzog:2012kx} with emphasis on the spin physics.

 {\bf Acknowledgements}
This work was supported by Mid-career Researcher Program through NRF grant No. 2010-0008456. 
 It is also partly supported by the National Research Foundation of Korea(NRF) grant  through the
SRC program CQUeST with grant number 2005-0049409.

\end{document}